# Reaction-kinetics of organo-clay hybrid films: in-situ IRRAS and AFM studies

Syed Arshad Hussain, Md N. Islam and D. Bhattacharjee

Department of Physics, Tripura University, Suryamaninagar-799130, Tripura, India.
Email: sa_h153@hotmail.com, tuphysic@sancharnet.in
Phone: +913812375317 (O)
Fax: +913812374802 (O)

**Abstract:**

In this paper we have reported the reaction kinetics of nano dimensional clay saponite and hectorite with an amphiphilic cation octadecyl rhodamine B (RhB) in hybrid Langmuir monolayer at the air-water interface. The surface pressure-molecular area ($\pi - A$) isotherms were strongly influenced by the presence of clay with the lift-off area of the cationic amphiphile shifted to progressively larger area. In-situ infrared reflection absorption spectroscopy (IRRAS) was used to demonstrate the reaction kinetics. Time taken to complete the reaction kinetics for RhB-hectorite hybrid films is larger than RhB-saponite hybrid films. Atomic force microscopic images of hybrid Langmuir-Blodgett films give compelling visual evidence of the incorporation of clay platelets into the hybrid films and density of which increases with the progress of reaction kinetics.

## 1. Introduction:

In-situ characterization of Langmuir monolayer at the air-water interface is always difficult because of the sensitivity limit imposed by a single monolayer [1]. Several techniques are employed to study the monolayer film on the water surface, such as $\pi - A$ isotherms, grazing incidence X-ray diffraction (GIXRD), Brewster angle microscopy (BAM) and Fluorescence imaging microscopy (FIM) [2-6]. It is known that thermodynamic measurements such as $\pi - A$ isotherms cannot reveal detailed microscopic information. BAM and fluorescence imaging microscopy can be employed to directly visualize the morphology of the monolayer composed of molecules. However, these observations are confined to macroscopic and mesoscopic scales [7] and can determine neither molecular characteristics such as conformation and packing of the alkyl chains nor structure and interaction pattern of the head groups. The GIXRD technique is a valuable tool to obtain direct structural information of crystalline films at the air-water interface on the sub nanometer scale [5,6], but is limited by the low scattering intensity arising from the monolayers at the interface [8]. Infrared reflection-absorption spectroscopy (IRRAS) has emerged as one of the leading methods for structural analyses of monolayer at the air-water interface over the last decade [9-14]. The IRRAS technique not only allows the characterization of chain conformation and headgroup structure but also provides qualitative / quantitative informations about molecular orientation.

Nanostructured organic – inorganic hybrid ultrathin films have gained widespread interest because of their potential use in a number of diverse technological applications. The organic materials impart flexibilities and versatility whereas, the inorganic clay particles provide mechanical strength and mechanical stability and can have unique conducting, semiconducting or dielectric properties. Construction of such organic-inorganic nanostructured materials is an important target of modern materials research. The research has been motivated by a purpose of developing functional materials such as sensors, electrode-modifiers, nonlinear optical devices and pyroelectric materials [15-17].

Clay minerals are inorganic sheet like particles that can be incorporated into ultrathin hybrid films [18, 19]. It is possible for organo-clay hybrid films to create new functionalized materials because they possess characteristics such as an easily changeable layer-by-layer structure or precisely controllable film thickness, or a variety of physical properties, which are not present in each of the separate components [20]. Certain physical properties such as second-order nonlinear optics and magnetics of these hybrid films only exist when a noncentrosymmetric molecular alignment and/or an oriented molecular alignment in the film is present. Also the organization and properties of organic cations adsorbed onto a clay particle at the surface of the sub phase is different from that of pure component at sub phase surface [20].

One of the outstanding properties of the clay is the simultaneous incorporation of polar or ionic molecules into the interlamellar spaces (intercalation) [21]. The property of intercalation makes it easy to prepare the composite materials. If the orientation of the incorporated molecules can be controlled, the clay composite materials would be applicable to devices for current rectifying, nonlinear optics, and one-way energy transfer [21, 22].

The Langmuir-Blodgett (LB) method has been used to fabricate ultrathin hybrid films consisting of layers of anionic phyllosilicate clay (smectite) and amphiphilic cations [23-25]. LB technique is an excellent method to construct ultrathin hybrid clay films since it operates at ambient condition and allows the control of thickness and structure in monolayer and multilayers. LB films generally possess a high degree of organization [1, 126]. When a dilute solution of the amphiphilic cation in chloroform is spread over a dilute aqueous clay dispersion in the LB trough, the clay particles get adsorbed onto the floating amphiphile layer forming a floating hybrid monolayer of the clay-amphilhile complex. In the present investigation, infrared reflection-absorption spectroscopy (IRRAS) technique has been used to study the reaction kinetics of hybrid clay films at the air-water interface. To the best of our knowledge the study of reaction kinetics of the hybrid clay films using IRRAS spectra have not been reported previously. Evidence of the formation of monolayers of the smectite-amphiphilie complex has included surface pressure molecular area ($\pi - A$) isotherm measurements, Atomic Force Microscopy (AFM) and IRRAS. The surface pressure-molecular area ($\pi - A$) isotherm measurements are strongly influenced by the presence of the clay with the lift-off area of the cationic amphiphile being shifted to progressively larger area.

## 2. Experimental:

### 2.1. Materials:

Octadecyl rhodamine B (RhB, 99%, Molecular Probe) was used as received. RhB was dissolved in HPLC grade chloroform (99.9 %, Aldrich). The clay minerals hectorite and saponite used in this study were obtained from the Source Clays Repository of the Clay Minerals Society. The clay minerals were $Na^+$ saturated by repeated exchange with 1 M NaCl solution and washing. The particle size fractions between 0.5 and 2 µm was obtained by centrifugation. The clay minerals were stored as freeze-dried powders.

### 2.2. Isotherm measurement and film formation:

A commercially available Langmuir-Blodgett (LB) film deposition instrument (APEX 2000C, India) was used for isotherm measurement and hybrid monolayer film preparation. Either Milli-Q water (resistivity 18.2 MΩ-cm) or clay dispersions stirred for 24 h (by a magnetic stirrer) in Milli-Q water was used as sub phase. The clay concentration was fixed at 10 mg/L (10 ppm). 30 micro-litre of chloroform solution of RhB (1mM) was spread on the air-clay dispersion interface of the LB trough by a micro syringe. When the clay particles come in contact with the floating RhB monolayer strong electrostatic interaction is occurred between the negatively charged clay particles and the positively charged RhB. Thus a hybrid Langmuir monolayer of RhB and clay is formed. Allowing 30 minutes to complete the hybrid film formation, the floating hybrid monolayer was compressed at a rate of 10 $cm^2 min^{-1}$ to monitor the pressure-area isotherm. Surface pressure was recorded using a Wilhelmy arrangement. Each isotherm was repeated several times and consistent results were obtained. To monitor the reaction kinetics of RhB and clay, the barrier compression was started with different waiting time after spreading the RhB molecules on the aqueous clay dispersion viz. 5 and 30 minutes. In both the cases the films were deposited in upstroke (lifting speed 5 mm $min^{-1}$) at a fixed surface pressure of 15 mN/m onto smooth silicon substrate for AFM measurements.

### 2.3. Infrared reflection absorption spectroscopy:

Infrared reflection absorption spectroscopy (IRRAS) of the Langmuir monolayer at the air-water interface was done by a special accessories attached with a Bruker IFS66v/s FTIR spectrometer. For IRRAS measurement the monolayer film on the liquid surface was prepared in a PTFE trough (area=17.6 $cm^2$) with a manually controlled barrier. A schematic representation of the IRRAS set-up [ref] for measurements of monolayer film on the liquid surface is shown in figure 1. The trough was filled with aqueous dispersion of either saponite or hectorite. A microsyringe was used to spread 3 $\mu l$ chloroform solutions (1mM) of RhB on the air-clay dispersion sub phase of the trough. Then the spectra were recorded with regular time interval. A total of 512 scans were signal averaged using an optical resolution of 2 $cm^{-1}$. The films on the liquid surface were measured at $40^0$ incidence to have maximum signal strength. The background spectra was the clay-dispersion surface before spreading the RhB solution. The spectrometer used in this study was equipped with a liquid nitrogen cooled MCT detector and a KBr beam splitter.

### 2.4. Atomic Force Microscope:

The Atomic force Microscope (AFM) images of RhB-clay hybrid monolayer films were taken in air with a commercial AFM system Autoprobe M5 (Veeco Instr.) using silicon cantilevers with a sharp, high apex ratio tip (UltraLevers™, Veeco Instr.). All the AFM images presented here were obtained in intermittent-contact ("tapping") mode. Typical scan areas were 3×3 $\mu m^2$. The monolayers on Si wafer substrates were used for the AFM measurements.

## 3. Results and discussions:

### 3.1. Surface pressure-molecular area ($\pi - A$) isotherm:

Figure 2 shows the surface pressure – area per molecule isotherms of RhB on pure water, 10 ppm saponite and 10 ppm hectorite sub phase.

The isotherm of RhB on water sub phase (curve a) shows a gradual rise with the surface pressure starting at a molecular area 1.7 $nm^2$ per molecule and collapses at 0.60 $nm^2$ per molecule (32mN/m). These values as well as the shape and nature of the RhB isotherm on pure water are consistent with the reported results [27, 28].

The isotherms of RhB on 10 ppm saponite (curve b) and 10 ppm hectorite (curve c) are also shown in figure 2. In presence of clay the isotherms are shifted towards larger area. For monolayers on clay dispersion sub phase the onset of the isotherms are 1.83 and 2.06 $nm^2$ and the surface pressures increase almost linearly up to 45.9 and 49.8 mN/m for saponite and hectorite respectively.

The relaxation curves of these monolayers are also shown in the inset of figure 2. The area loss of the RhB isotherm on pure water relaxed at 15 mN/m after 85 min are close to 20%. Where as the area loss for RhB monolayer in presence of clay decreases to 16% and 5% for hectorite and saponite respectively. This is an indication of more stable film in presence of clay platelets. Stability of the monolayer film increases with increase in collapse pressure [1]. In presence of clay platelets collapse pressure increases to a larger extent than that in absence of clay platelets. This is also an indication of increase of stability of the monolayer film in presence of clay platelets.

In this context it is important to mention that electrostatic interactions between cationic dyes and negatively charged clay platelets are occurred at the air-water interface. This electrostatic interaction can also be described as an ion exchange reaction, as the charge compensating cations of the smectites are exchanged by the positively charged dye molecules [29]. Thus the clay particles come at the air-water interface and incorporated into Langmuir monolayer. Consequently it affects the monolayer characteristics and increases the stability of the monolayer films.

#### 3.1.1. Apparent compressibility of the hybrid monolayer:

To obtain more information about the monolayer films, the compressibility of the monolayer films have been calculated from the isotherm characteristics. The apparent compressibility, C, is defined as

$$C = -\frac{1}{a_1}\frac{a_2 - a_1}{\pi_2 - \pi_1}$$

Where $a_1$ and $a_2$ are the areas per molecule at surface pressures $\pi_1$ and $\pi_2$ respectively [30, 31]. In the present case $\pi_1$ and $\pi_2$ are chosen as 10 and 30 mN/m surface pressures respectively. The compressibility of RhB monolayer on pure water is 24.1 $mN^{-1}$. For RhB-saponite monolayer the compressibility value is 19.7 $mN^{-1}$ and for RhB-hectorite is 23.6 $mN^{-1}$. This suggests that the monolayer of RhB on pure water is more compressible than that of the monolayer of RhB in presence of clay platelets which are thought to make the films hard. Judging from the C values, however, the monolayers of RhB-hectorite are comparatively softer in comparison with those of RhB-saponite monolayer.

### 3.2. Infrared Reflection Absorption Spectroscopy:

The IRRAS technique is especially attractive because it can be applied to obtain spectra from hybrid clay-organic monolayers directly on the air-water interface. The infrared reflection absorption spectra of RhB monolayer on 10 ppm saponite and hectorite dispersion sub phases measured at different time interval after spreading are shown in figures 3a and b respectively. In the figures the $\nu$(Si-O) region of the spectra are presented. It was observed that the absorption bands are negative [figure not shown], which is typical for IRRAS at dielectric medium [32]. Here in the present figures we have presented the transmittance so that the bands are positive.

The IRRAS spectra possess strong prominent band at 996 $cm^{-1}$ (figure 3a). This 996 $cm^{-1}$ band corresponds to the in-plane Si–O stretching vibration, $\nu$(Si–O), of saponite [33]. As the waiting time after spreading increases, the reflectance becomes more intense and reaches it's maximum with waiting time 15 minutes. When the reflectance at 996 $cm^{-1}$ is plotted as a function of the waiting time (inset of figure 3a), a nonlinear relationship is observed and after 15 minutes the reflectance becomes almost constant. This is an indication of completion of reaction kinetics of clay with RhB at the air-water interface. When the negatively charged clay platelets come in contact with the amphiphilic RhB cation at the air-water interface, nano order clay platelets are adsorbed at the bottom of the floating monolayer of RhB cation by electrostatic interaction. This interaction takes some time and number of clay platelets adsorbed onto the floating layer increases with time resulting in an increase in the thickness of the hybrid monolayer at the air-clay dispersion interface. This increase in thickness causes an increase in reflectance of 996 $cm^{-1}$ band. However after a certain time interval all the floating amphiphilic cations are neutralized by the clay particles and no free RhB cation exists for further adsorption of clay platelets, resulting in the attainment of maximum thickness of RhB-clay hybrid monolayer. Consequently the reflectance also reaches its maximum and becomes almost constant. The constant reflectance after waiting time 15 minutes is an indication of the end of the interaction.

The $\nu$(Si-O) vibration at 1001 $cm^{-1}$ of the hybrid RhB-hectorite is also due to in-plane orientation of the Si-O group. Here the reflectance increases gradually up to 20 minutes and a rapid change of reflectance have been observed in between 20 - 25 minutes after which the intensity of 1001 $cm^{-1}$ band becomes constant indicating completion of reaction kinetics between the negatively charged hectorite and positively charged RhB amphiphile at the air-water interface.

It is worthwhile to mention in this context that the time taken for the completion of reaction kinetics between clay and amphiphilic cation RhB at the air-water interface is different for saponite and hectorite. For saponite it is 15 minutes whereas for hectorite this value is 25 minutes. Although for both the case same amphiphilic cation RhB was used however this difference in reaction time is mainly due to the difference in shape and cation exchange capacity (CEC) of the two different clay saponite and hectorite. Similarly if an amphiphilic cation other than RhB is used then the reaction time may vary depending on the size, shape and charge of the cation.

Schematic diagrams of organization of RhB-clay hybrid films at the air-clay dispersion sub phase are shown in figure 4. Immediately after spreading few clay particles come in contact with the floating RhB layer at the air-water interface and form complex (figure 4a). However there exists lot of free amphiphile cations for further reaction to occur. With passage of time more and more clay particles come in contact with the floating RhB molecules and form complex. After a certain time interval all floating amphiphiles are neutralized by clay particles indicating the completion of reaction kinetics (figure 4b). At that point the intensity of $\nu$(Si–O) band becomes constant (figure 3a & b). After this time interval there exists no free amphiphilic cation for further reaction to occur. AFM observations strongly support this explanation.

### 3.3. AFM Observation of Monolayer Films:

The morphology and surface structure of the RhB-clay monolayer films transferred onto solid substrate were studied by AFM. Figures 5a & b show the AFM images of the hybrid films when the monolayer of RhB on clay suspension deposited onto smooth silicon substrate at 15 mN/m with waiting time (i) 5 minutes and (ii) 30 minutes after spreading the RhB solution on 10 ppm saponite and 10 ppm hectorite suspensions respectively. In the AFM images of the films when waiting time was 5 minute, all clay (both saponite & hectorite) particles were separated from each other [figures (i) of 5(a) and 5(b)]. Since the film exhibited definite surface pressure, the space between the clay particles was concluded to be occupied with RhB cations. From the height profile analysis it is seen that the thickness of the films lies in between 1 to 2 nm for all the films. However, when the waiting time before compression was 30 minutes instead of 5 minutes the clay particles are in close contact with little vacant space between the particles. When the waiting time was 30 minutes the whole surface of the film was covered with the clay particles. The clay particles were packed more densely in RhB-hectorite hybrid film than that of RhB-saponite hybrid film and partially overlap. Increase in surface coverage of the monolayer films with increasing waiting time after spreading the amphiphilic cation before compression is an indication of completion of reaction kinetics of nano-clay platelets and RhB at the air-water interface. These AFM images reveal that the clay particles hybridize RhB cations at the air-water interface and gives compelling visual evidence of reaction kinetics and incorporation of both saponite and hectorite clay platelets onto the monolayer dye-clay hybrid films.

### 3.4. Conclusion:

In this paper we have demonstrated the reaction kinetics of nano dimensional clay saponite and hectorite with an amphiphilic cation octadecyl rhodamine B (RhB) in hybrid Langmuir monolayer at air-clay dispersion interface. The surface pressure-molecular area ($\pi - A$) isotherm measurements were strongly influenced by the presence of the clay with the lift-off area of the cationic amphiphile being shifted to progressively larger area. Inclusion of clay particles increases the stability of the monolayer films. RhB-hectorite film was softer than RhB-saponite hybrid film. In-situ IRRAS reveals that time taken to complete the reaction kinetics for RhB-saponite hybrid films was 15 minutes where as for RhB-hectorite hybrid film this value was 20 minutes. This reaction time may vary if a different amphiphilic cation is chosen depending on the shape, size and charge of the cation. Atomic force microscopic images of hybrid Langmuir-Blodgett films give compelling visual evidence of the incorporation of clay platelets into the LB film and increase in density of clay platelets with the progress of time is an visual evidence of reaction kinetics.

### Acknowledgement:

DB and SAH are grateful to CSIR, Govt. of India for financial assistance (Ref. No.03 (1080)/06/EMR-II) to carryout this research work.

**Figure caption:**

Figure 1: Schematic representation of the IRRAS set-up for measurements at the air-liquid interface.

Figure 2: Surface pressure – area per molecule ($\pi - A$) isotherm of RhB on (a) pure water, (b) 10 ppm saponite and (c) 10 ppm hectorite sub phase. Inset shows the stability curves of RhB monolayer on (a) pure water, (b) 10 ppm saponite and (c) 10 ppm hectorite sub phase.

Figure 3a: Infrared reflection absorption spectra (IRRAS) of RhB surfactant monolayer on 10 ppm saponite clay dispersion sub phase measured with waiting time 5, 10 and 25 minutes after spreading. Inset shows the plot of $\nu$(Si-O) band intensity as a function of time.

Figure3b: Infrared reflection absorption spectra (IRRAS) of RhB surfactant monolayer on 10 ppm hectorite clay dispersion sub phase measured with waiting time 3, 21 and 27 minutes after spreading. Inset shows the plot of $\nu$(Si-O) band intensity as a function of time.

Figure 4: Schematic representation of RhB-clay hybrid films at air-water interface with (a) immediately after spreading of RhB cation onto the clay dispersion and (b) after completion of reaction kinetics ie after the completion of hybrid film formation when all the RhB cation are neutralized by anionic clay at the air-water interface.

Figure 5a: AFM image of monolayer RhB-saponite LB film deposited onto a smooth silicon substrate at 15 mN/m with waiting time (i) 5 and (ii) 30 minute before compression begins.

Figure 5b: AFM image of monolayer RhB-hectorite LB film deposited onto a smooth silicon substrate at 15 mN/m with waiting time (i) 5 and (ii) 30 minute before compression begins.

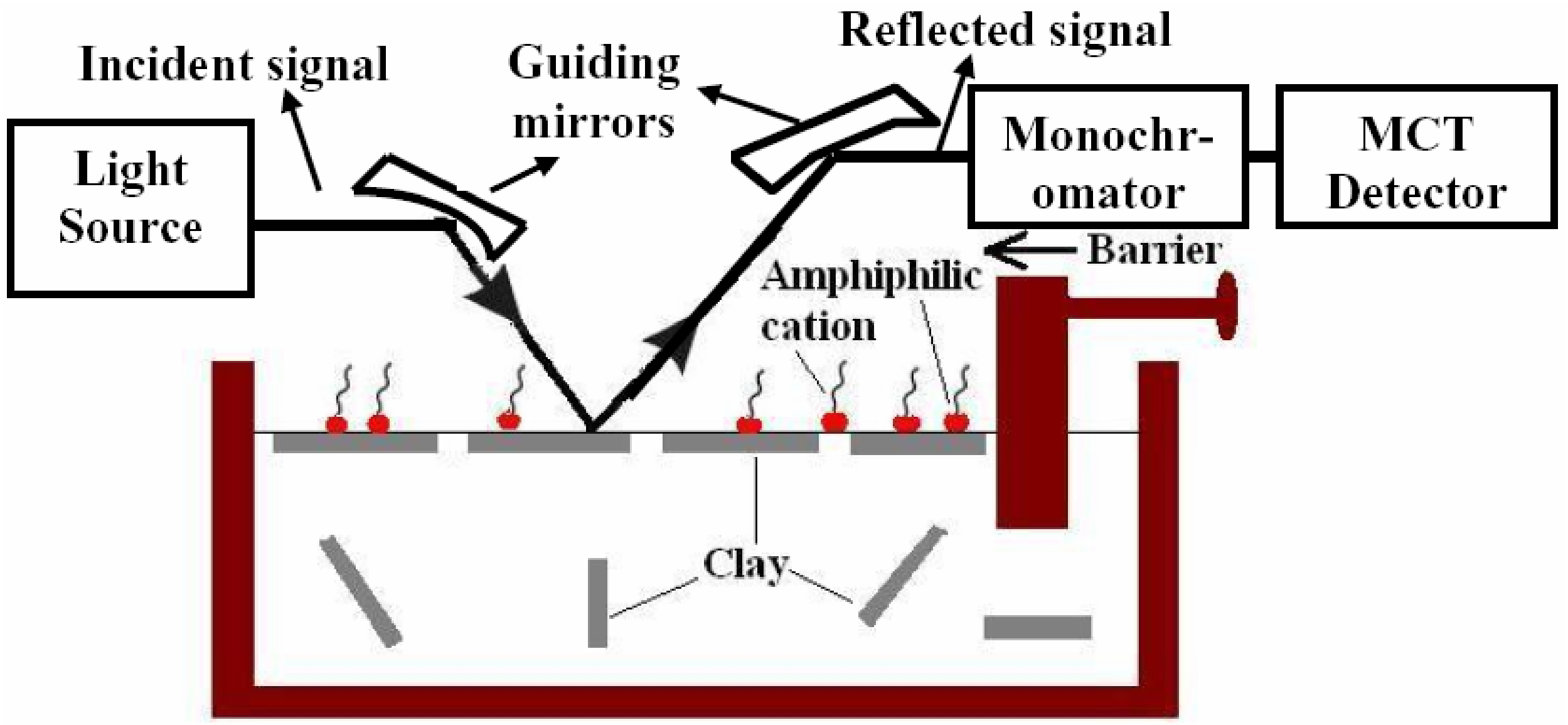

Figure 1: Syed Arshad Hussain et. al.

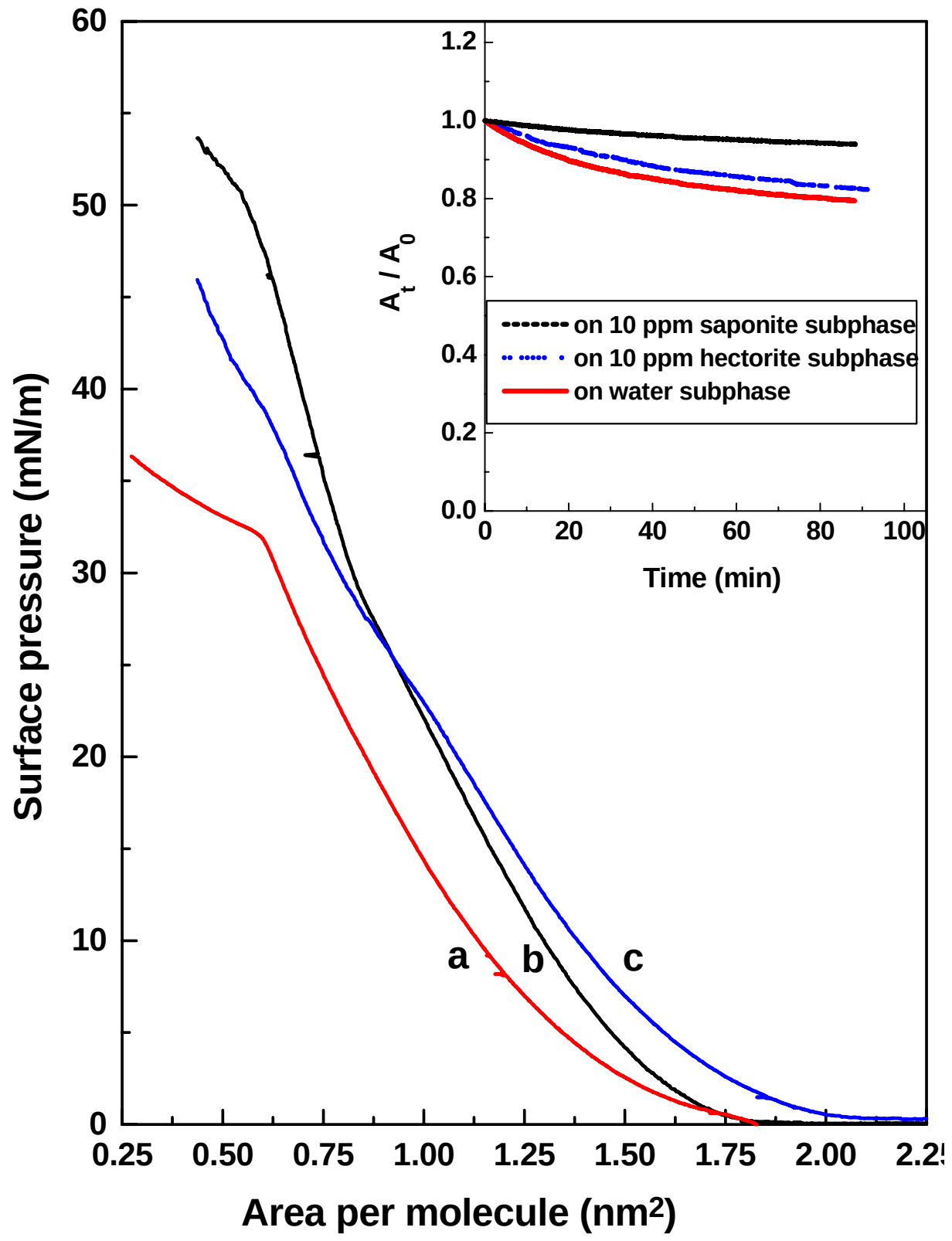

Figure 2 Syed Arshad Hussain et. al.

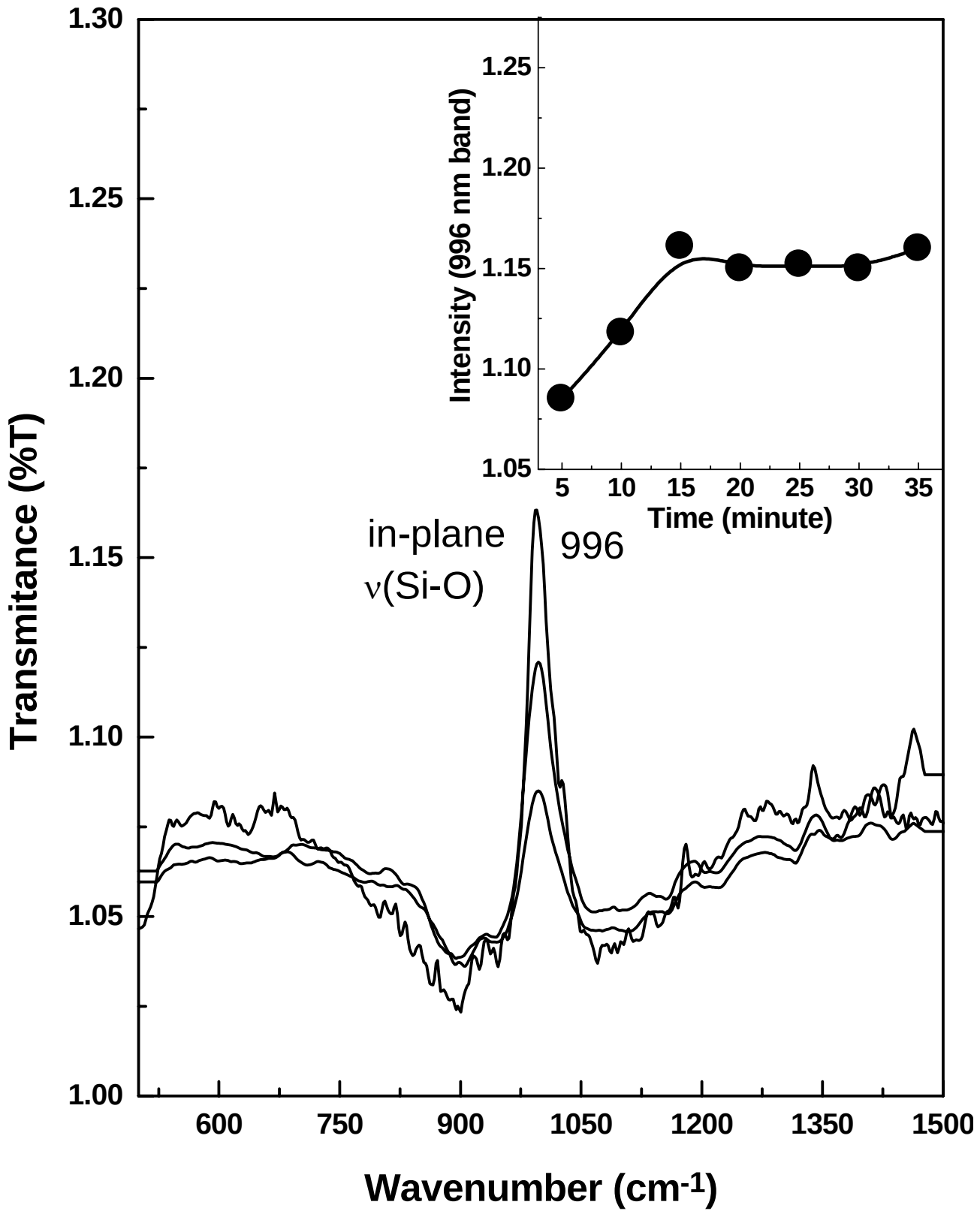

Figure 3a Syed Arshad Hussain et. al.

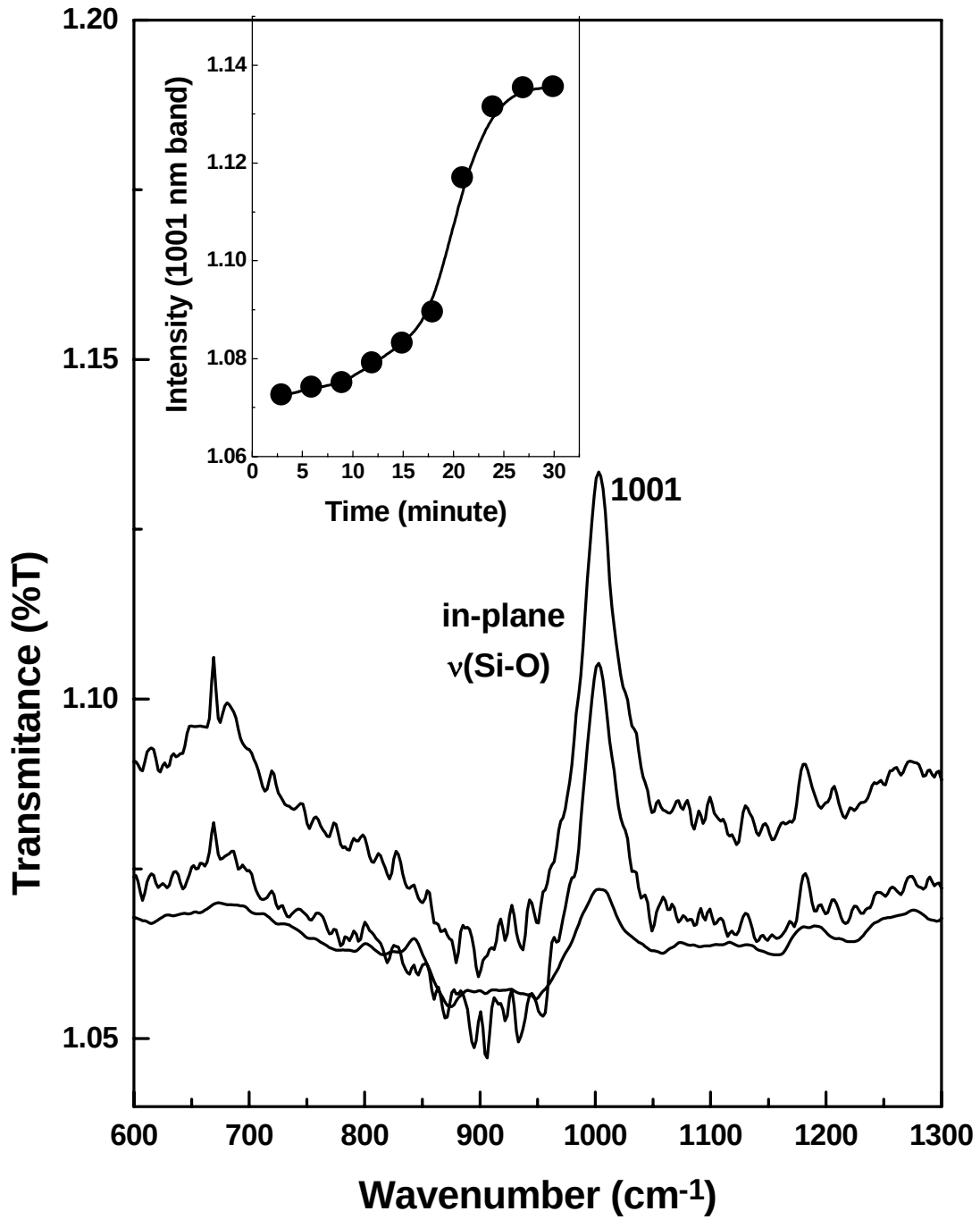

Figure 3b Syed Arshad Hussain et. al.

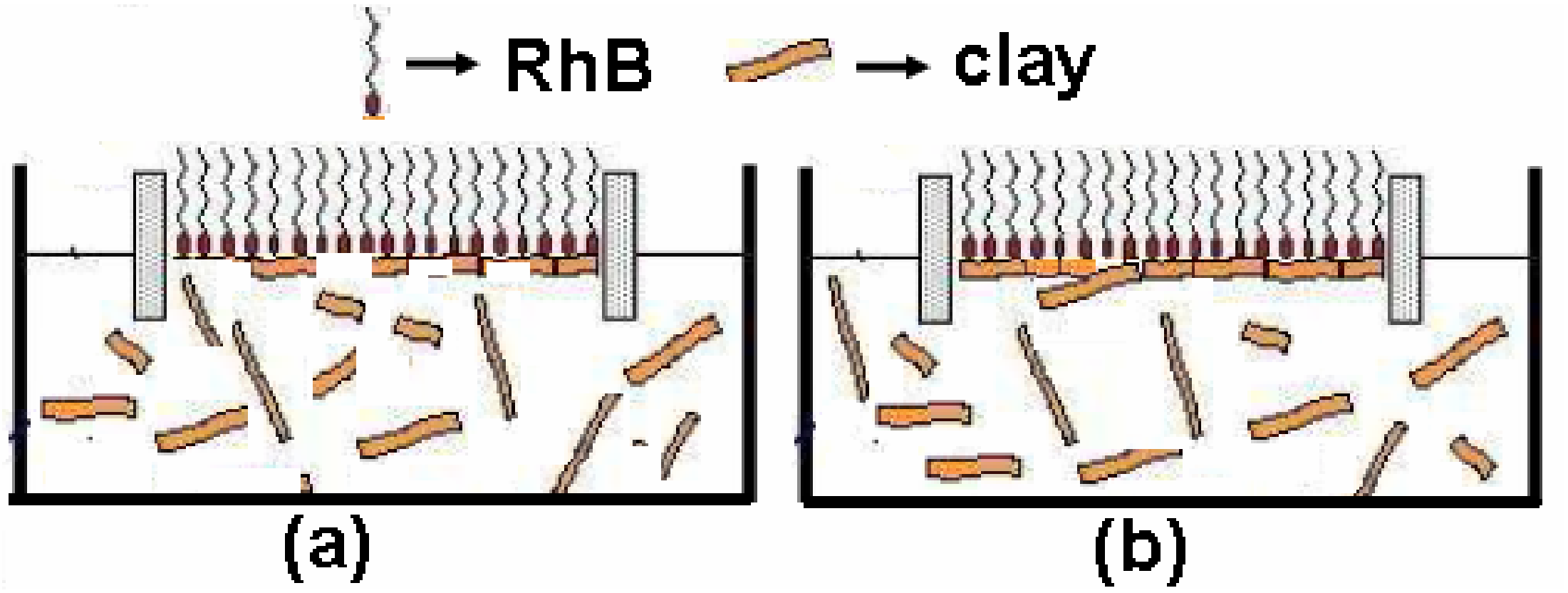

Figure 4 Syed Arshad Hussain et. al.

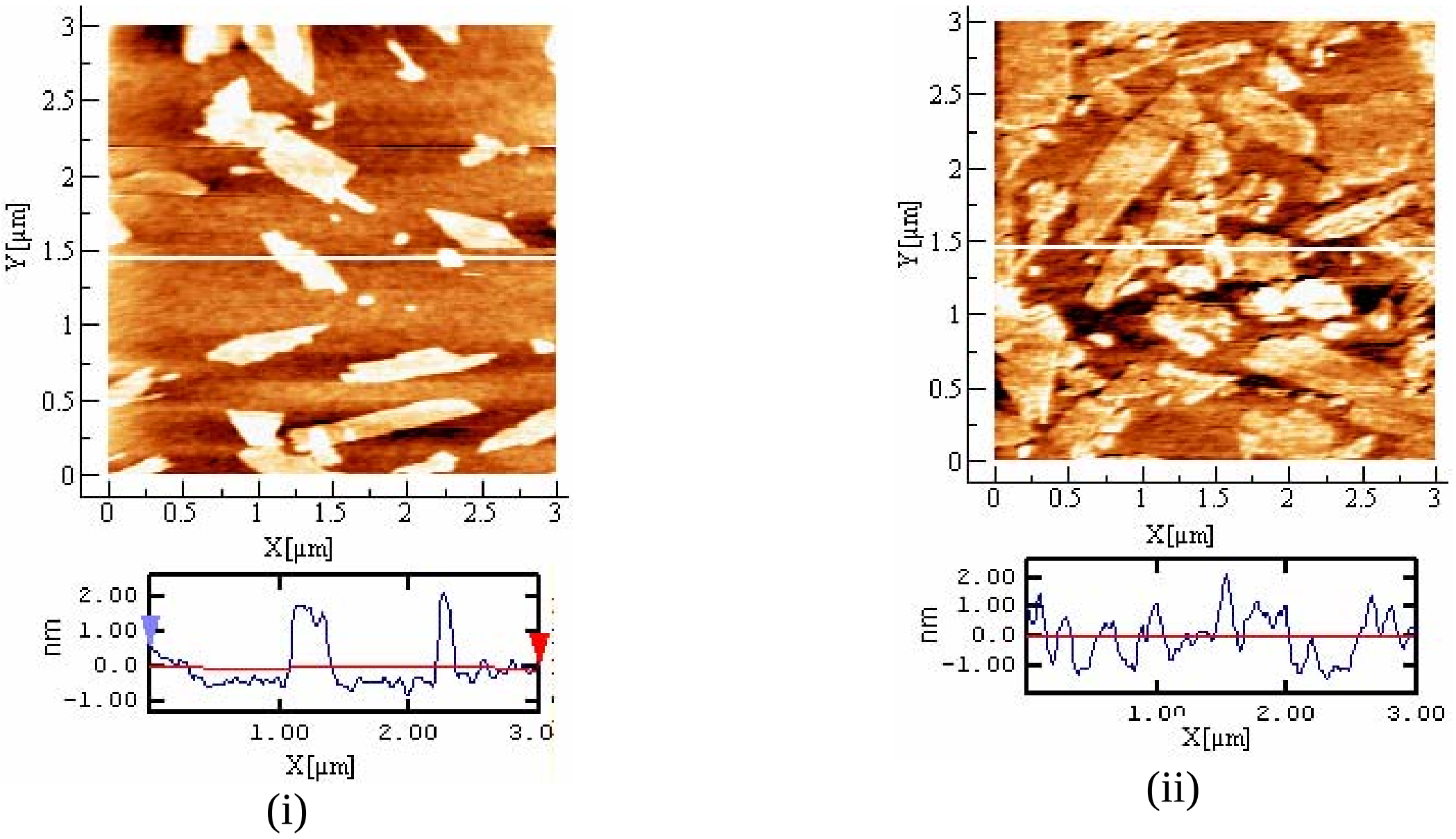

Figure 5a Syed Arshad Hussain et. al.

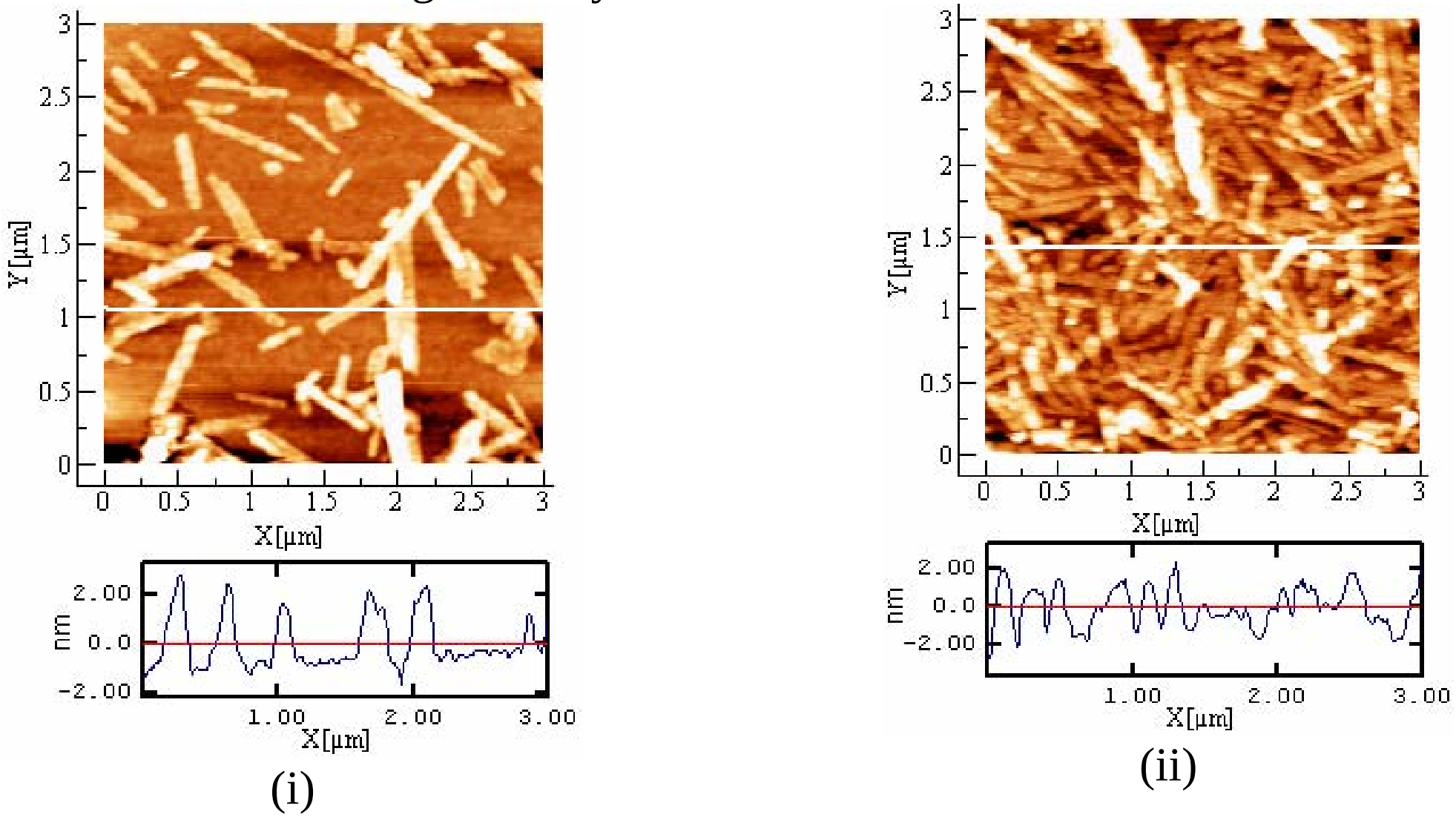

Figure 5b Syed Arshad Hussain et. al.